\documentclass[doublecol]{epl2} 
\usepackage[utf8]{inputenc}
\usepackage[T1]{fontenc}
\usepackage{graphicx}
\usepackage{dcolumn}
\usepackage{stfloats}
\usepackage{lipsum}
\usepackage{xcolor}
\usepackage{amsmath}
\newcommand{\ket}[1]{\left| #1 \right\rangle}
\newcommand{\bra}[1]{\left\langle #1 \right|}
\newcommand{\op}[1]{#1}

\begin{document}

{}
{}
{}

\newcommand{\change}[1]{\textcolor{blue}{#1}}

\title{\revision{Driven emergent phases in small interacting condensed matter systems}}

\author{
  Gayanath W. Fernando\inst{1} and 
  R. Matthias Geilhufe
  \inst{2}, and
  Adil-Gerai Kussow\inst{1},  and
  W. Wasanthi P. De Silva\inst{3}}

\institute{                    
  \inst{1} 
  University of Connecticut, 196A Auditorium Road, U-3046 Storrs, CT 06269-3046, USA\\
  \inst{2}
  Nordita, KTH Royal Institute of Technology and Stockholm University, Roslagstullsbacken 23, SE-106 91 Stockholm, Sweden\\
  \inst{3}
 Department of Physics, University of Sri Jayewardenepura, Nugegoda, Sri Lanka}

\abstract{
 Single- and many-electron calculations and related dynamics are presented for a dimer and small Hubbard
clusters. Floquet-Bloch picture for a periodic dimer is discussed  with regard to the time dependence of the Peierls gap and the expectation of the current operator.
In driven Fermi-Hubbard clusters, the time dependence of
Charge Gaps and phase separation along with charge pairing at various cluster sizes indicate the presence and absence of paired electron states.  We examine the effect of electromagnetic time dependent external perturbations on Hubbard many-electron systems in our search of for precursors to superconducting states and time crystals. Two principally different kinds of electromagnetic excitations are analyzed:(1) Recently demonstrated dynamic modulation of Hubbard parameters due to excitation of certain phonon modes within the far-infrared domain, and (2) Hubbard Hamiltonian, with fixed parameters in an electromagnetic field, resonant with transitions between the ground state and high-energy excited states as possible precursors to superconductivity, within visible-near-infrared domains.}

\maketitle

\section{Introduction}
 During the past decade, our understanding of periodically driven
   systems has advanced significantly for example, in cold atoms, electron-phonon and strongly correlated systems.
 It is now possible to use fast, time dependent drivers, such as lasers, to dynamically
 induce phase/topological changes in the electronic properties of a given material.
 Some of the recently developed ultra-fast
 techniques have been able to drive condensed matter systems out of equilibrium into transient novel phases. Such phenomena are
related to symmetry breaking, a well known concept of modern physics.
   Strongly correlated materials give rise to a wide variety of interesting and exotic phenomena leading to complex phase diagrams and dynamical responses.
    Recent reports of photo-molecular high temperature  superconductivity~\cite{Cavalleri}, transient dehybridization in a $f$-electron system~\cite{Leuen} and photo-induced changes in the second harmonic generation signal from a transition metal oxide~\cite{Sheu} are some examples of correlated electron systems
 where emergent phases could exist under certain conditions.
 Recently, breaking of time translation symmetry has been observed in Floquet systems,
 for example, with emergence of a "time crystal" through persistent
 oscillations of interacting spins~\cite{zhang}.
 
 The unifying themes of this manuscript are stable dynamical phases and their properties in simple systems, focusing on dynamical tuning of one-and many-particle gaps. In a periodic dimer where one-particle physics is central, the ability to open or close gaps has implications in controlling topological and transport properties while in small Hubbard clusters, tuning many-electron gaps can lead to non-topological dynamical phase separation.
  We also examine the dynamics of paired electron states and their direct and indirect enhancements, possibly related to  unconventional superconductivity. 
 For example, charge pairing (or electronic phase separation) instabilities could produce conditions favorable to superconductivity, acting as a precursor. In addition, several dynamical effects arising from external time dependent sources (periodically driven and/or photo-induced) could also be favorable in this regard and are discussed here.
   
Certain electronic inhomogeneities found in  complex oxides~\cite{dagotto} are regarded as resulting from such strong correlations and could be labeled as electronic phase separation. Here, when doping the half-filled antiferromagnetic insulator with holes, the energy of the system becomes lower if the holes cluster to create variations of hole rich and hole poor regions due to antiferromagnetism. Our group was able to  point out various pairing scenarios using the Hubbard model based on several different bipartite clusters\cite{PRB2007,PRA2012} using exact calculations. 
These results show similar phase separation at larger doping levels even around optimal doping at the cluster level, but the cluster calculations are invariably tied to some uncertainties due to  size and edge effects. However, with the progress of ultra-cold atom traps and the ability to generate and control small systems, cluster calculations (using Bose-Hubbard or Fermi-Hubbard-type Hamiltonians) can be expected to provide useful guidance.

\section{Dynamical Symmetry Breaking and Floquet Picture}
Periodically driven quantum matter can be described in the Floquet-Bloch formalism, based on a Hamiltonian which is invariant under discrete lattice translations ${\bf R}_i$ and time translations $ T$ (the period of the driver), 
$H({\bf x} +{\bf R}_i, t+T) = H({\bf x}, t)$ where $T=2\pi/\omega.$ Here $\omega$ denotes the frequency of the driver. Using the Floquet-Bloch ansatz, we can express these states as $\ket{\Psi_{\alpha \vec{k}}} = e^{\mathrm{i}\left(\vec{k}\cdot\vec{r}-\epsilon_{\alpha,\vec{k}}t\right)}\left|u_{\alpha,\vec{k}}(\vec{r},t)\right>$, where $\vec{k}$ is the Bloch wave vector, $\epsilon_{\alpha,\vec{k}}$ the quasi-energy, and $\alpha$ a band index. $\left|u_{\alpha,\vec{k}}(\vec{r},t)\right>$ are time and lattice periodic states and tied to Floquet-Bloch creation and annihilation operators $\op{c}^\dagger_{\alpha\vec{k}}(t)$ and $\op{c}_{\alpha\vec{k}}(t)$, as well as their Fourier components $\op{c}^\dagger_{\alpha\vec{k}n}$ and $\op{c}_{\alpha\vec{k}n}$, e.g., $\op{c}_{\alpha\vec{k}}(t) = \sum_n e^{\mathrm{i} n\omega t} \op{c}_{\alpha\vec{k}n}$. The time-dependent Schr{\"o}dinger equation can now be mapped to a time-independent eigenvalue problem
   \begin{equation}
       \sum_{m \beta} (H_{\vec{k}\,\alpha\beta}^{n-m} - m \omega\delta_{mn} \delta_{\alpha\beta}) \ket{\Phi_{\beta\vec{k}}^m}=\epsilon_{\alpha}\ket{\Phi_{\alpha\vec{k}}^{n}},
       \label{eigv}
   \end{equation}
   with $\ket{\Phi_{\alpha\vec{k}}^{n}} =  \op{c}^\dagger_{\alpha\vec{k}n} \ket{0}$. Note that in practice, only a finite number of Fourier components is taken into account.
  
   \begin{figure}
\includegraphics[width=16pc]{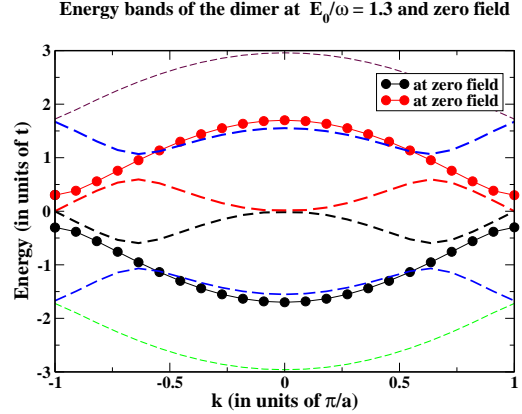}
\caption{
 Floquet band structure for the dimer discussed here.
Note that the zero field bands show the typical Peierls gap at the zone boundaries for the parameters chosen (see text).
Also note the time-dependent floquet bands (dashed lines) that close the standard Peierls gap in the dimer at the zone edges at a field strength of $E_0/\omega = 1.3$. The field strength has a significant effect on the floquet band dispersions.}
\label{floquet_bands}
   \end{figure}   
   
   \begin{figure}
	   \includegraphics[width=16pc]{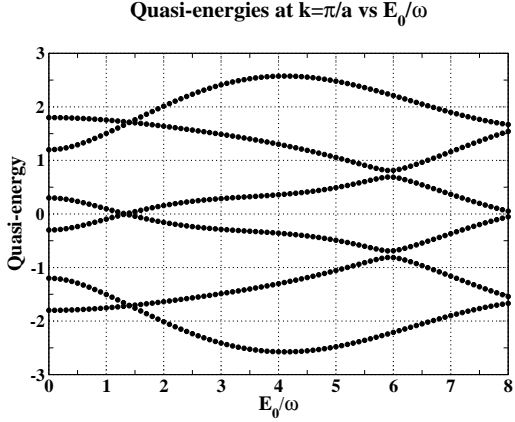}

\caption{ Floquet quasi-energies vs the field strength: Note that at $E_0/\omega = 1.3$ there are level crossings that lead to
closing the Peierls gap and transitions discussed in the text. The Fourier modes corresponding to $n, m=-1,0,1$ are included in the expansion
in Eq.~\ref{ham_time}.}
\label{floquet}
\end{figure}

Recently, G{\'o}mez-Le{\'o}n and
Platero~\cite{gomez} have studied a one-dimensional, spatially periodic dimer under a time-periodic vector potential
 $ A(t) = A_0 \sin(\omega t)$ with $A_0=qE_0/\omega$ and time dependent hopping $\tau$ (with $E_0$ denoting the field strength). The corresponding time-dependent Hamiltonian at a given {\bf k}-point may be written as
   \begin{equation}
	   H_{\bf k}(t)= \sum_{n,m}\sum_{j,l}\tau(t)_{j,l}c^{\dagger}_{{\bf k},n}c_{{\bf k},m}
	   \exp(i\kappa(t)\cdot(\rho_{n,j}-\rho_{m,l}))
	   \label{ham_time}
   \end{equation}
   with $\kappa(t)=(-t,{\bf k})$ and $\rho_{n,j} = (n\omega,{\bf R}_j).$ Fourier transforms of the time-dependent hopping $\tau(t)$ are related to field induced states and this makes it possible to induce various transitions.
 \begin{figure}
    \centering
    \includegraphics{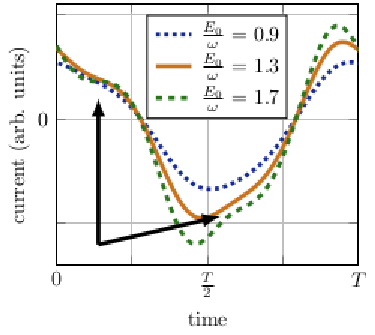}
    \caption{Estimated current for different field strengths. Beyond the critical field strength of $E_0/\omega=1.3$, higher harmonics in the current become dominant. This value coincides with the quasi-energy level crossing shown in Fig. \ref{floquet}.}
    \label{dimer:current}
\end{figure}  
   Here the time-dependent vector potential ${\bf A}(t)$ is included via the canonical substitution $\vec{k} \rightarrow \vec{k} + e \vec{A}(t)$. Hence, it is combined with exponentials $\exp(im\omega t)$ in a Fourier series expansion to absorb the time dependence into the Hamiltonian.
   Following time averaging,
   the resulting $\omega$-term in the above equation is analogous to an effective electric field applied along the extra dimension. Some of the emerging quasi-energies (defined modulo $\omega$) can also be thought of as being tied to higher topological orders. Non adiabatic processes in such systems increase the tuning possibilities with the addition of new topological phases.

\section{Periodic Dimer and Peierls Gap}
   As a simple example, we focus on the one-dimensional periodic dimer of atoms
   and calculate the time dependent
   effects on the band structure.
   The expected ground state in the time independent case
   corresponds to atoms with equal spacing. However, this arrangement is unstable and the well-known
   Peierls transition~\cite{ashcroft} results in
   dimerization or the formation of short and long bonds (alternate pairs of atoms moving closer).
   The undriven tight binding model gives rise to an energy band structure

   \begin{equation}
       E_{\pm}(k)=\pm t^{''}\sqrt
       {\lambda^2 + 1 + 2\lambda\cos(ka)}
   \end{equation}
   in the basis of two atoms with $\lambda=t^{''}/t^{'}$ being the ratio of hopping parameters corresponding to long and short bonds and $a$ the lattice translation distance
   (see Fig.~\ref{floquet_bands}).
   
   In this static case, the dimerization leads to an energy gap (Peierls gap) in the electronic bands
   at the edges of the Brillouin Zone
    which lowers  the total energy.
    This can be affected by a time dependent external field by inducing metastable states in the
    gap region. Figs.~\ref{floquet_bands} and \ref{floquet} show calculated band structure and quasi-energies for the driven case using parameters
    $\lambda=0.7$ and driving frequency $\omega = 1.5$ in units of hopping $t^{'}$. Note that the Floquet bands at a field strength of $E_0/\omega = 1.3$ close the 
    Peierls gap at the zone boundaries due to crossings seen in the quasi-energies at this field strength (Fig.~\ref{floquet}). 

   To shed light into the gap closing, we calculate the current expectation value
   \begin{equation}
       \left<\op{J}(t)\right> = \operatorname{Tr} (\rho(t) \op{J}(t)).
   \end{equation}
   The current operator is obtained as $\op{J}(t) = e \frac{\partial \op{H}_{k}(t)}{\partial k}$, taking into account the Hamiltonian of \eqref{ham_time}. The density matrix $\rho(t)$ is calculated  by composing it into $k$-dependent terms $\rho = \sum_k \rho_k$ and numerically solving the following Lindblad equation
   \begin{multline}
       \frac{\mathrm{d}\rho_k}{\mathrm{d}t} = -\mathrm{i} \left[\op{H}_k(t),\rho_k\right]+ \gamma \sum_{k'}\left(\op{L}_{k,k'}\rho_{k'}\op{L}_{k',k}^\dagger \right. \\ \left.
       - \frac{1}{2}\left\{\op{L}_{k',k}\op{L}_{k',k}^\dagger,\rho_k\right\}\right).
       \label{lindblad}
   \end{multline}
   The initial conditions are taken corresponding to the undriven system as follows:
   \begin{equation}
       \rho_k(0) = \left. \rho_{k,0}  \right|_{A_0 = 0} = \sum_{n=\pm} f(E_{k\,n},\mu,\beta) \ket{k n}\bra{k n},
   \end{equation}
   where $f(E_{k\,n},\mu,\beta)$ denotes the Fermi-Dirac distribution function for a state with energy $E_{k\,n}$ ($n=\pm$), $\mu$ the chemical potential, $\beta=(k_B T)^{-1}$. Furthermore, $\ket{k \pm}$ are eigenstates of the undriven Hamiltonian $\left. \op{H}_k \right|_{A_0 = 0}$. Here, $\ket{k\,-}$ ($\ket{k\,+}$) denote the occupied (unoccupied) states at $\mu=0$. \revision{The Lindblad jump operators are constructed as follows
   \begin{equation}
       \op{L}_{k'k} = \ket{k'\,-}\bra{k\,+}. 
   \end{equation}
   Note that we incorporate both, inter band transitions ($\ket{k\,+} \rightarrow \ket{k\,-}$) and intra band conditions ($\ket{k'\,\pm} \rightarrow \ket{k\,\pm}$). The inter band transitions ensure the system to relax into the ground state, whenever the driving-field is turned off. Intra band transitions induce a constant current in the system (offset in Fig. \ref{dimer:current})}.
   After long enough driving, the current develops a regular pattern shown in Fig. \ref{dimer:current} for one driving period. For a critical field strength of $E_0/\omega = 1.3$, the current develops a dominant feature related to higher harmonics. The value of the critical field strength coincides with the level crossing of the Floquet quasi-energies shown in Fig. \ref{floquet}. \revision{Parameters are chosen similarly as before, with a dissipation of $\gamma=0.1$ for inter band ($k=k'$) and $0.01$ for intra band transitions.} A Fourier analysis shows that the current responds in odd multiples of the driving frequency, ($\omega, 3 \omega, \dots$). 
    It is very likely that the kinks/distortions around the extremal points that are indicated by arrows in the steady state
 current plot (Fig.~\ref{dimer:current}) at and beyond a critical value of the field strength are due to quantum interference-related effects due to quasi-energy levels in the vicinity of a given crossing
 (as suggested in Ref.~\cite{russo}).

\begin{figure}
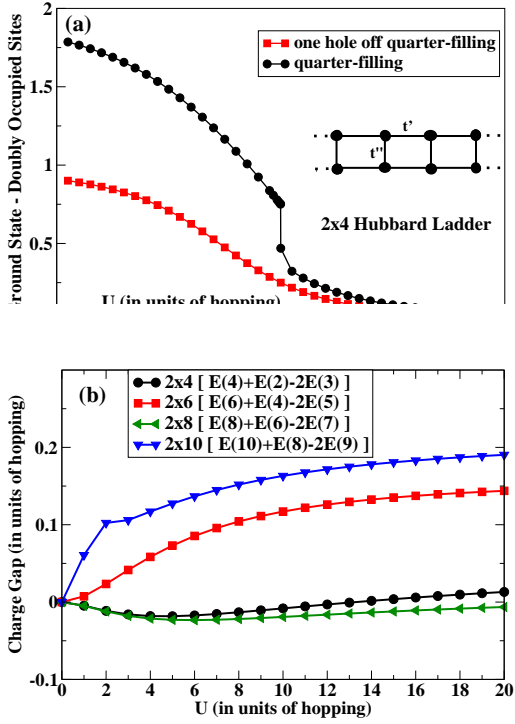

\includegraphics[width=16pc]{docc_full.eps}

\includegraphics[width=16pc]{ladder.eps}

\caption{(a) Doubly occupied sites in the ground state vs $U$ for the 2$\times$4 Hubbard ladder at two fillings: Note that at large negative $U$ values, the curves at given fillings approach their maximum allowed double occupancies (2 for quarter-filling and 1 for one hole off quarter-filling).
Also note that these double occupancies are fairly low compared to their allowed maximum values for positive $U$; the inset shows a typical 2$\times$4 ladder with indicated hopping parameters and periodic boundary
conditions along its axis.
(b) Charge Gaps for various 2$\times$N ladders at the indicated fillings, around one hole off quarter-filling. Note how the stability of the odd electron state vs paired combination varies with N.
\label{docc_full}}
\end{figure}

\section{Paired Electron States}

    Following the study of a one-particle gap closure in the dimer, we now focus our attention on the driven
    many-electron gaps and related properties in small Hubbard clusters.
    Minimal Hubbard cluster calculations for small ladders near half-filling have clearly shown a tendency to phase separate~\cite{PRB2007,PRA2012} leading to pairing. Here we examine the dynamical behavior of such clusters at various fillings for similar systems in order to examine the possibility of novel dynamical behavior.   For small clusters, we use exact diagonalization (ED) and Lanczos~\cite{Lanczos} method to calculate the eigenvalues and eigenstates of the single orbital (Fermi) Hubbard Hamiltonian $\mathcal{H}$ with periodic or open boundary conditions: 
    \begin{equation}
       \mathcal{H} = -\sum_{<ij>\sigma} t_{ij}B_{ij\sigma}  + 
       \sum_{i} U {n}_{i\uparrow}{n}_{i\downarrow}  
       \label{hamil_pulse}
\end{equation}  
  Here $U$ is the onsite Coulomb repulsion while $t_{ij}$ is the hopping matrix
   element from site $i$ to site $j$ and $\sigma$ denotes the spin index with
   $ B_{ij\sigma} ={c}^\dagger_{i\sigma}{c}_{j\sigma} + H. c.$ In what follows, we have
   considered nearest neighbor hopping (set to $t^{'}$) and if indicated, next nearest
   neighbor hopping (denoted by $t^{''}$). Note: We reserve the symbol $t$ to denote time throughout the manuscript. In addition, we have chosen to study the dynamics in the single orbital model due to its simplicity and relevance to the cuprates, while acknowledging that a multi-orbital model would be more appropriate to other classes of high temperature superconductors.

   Before including time dependent parameters in the above Hamiltonian, we focus first on the static case
   observables such as Charge Gap and double occupancies  defined using the many-body
   eigenvalues of $\mathcal{H}$ (note that we have only included the onsite
   Coulomb repulsion here but off-site repulsions can be incorporated easily).
   The Charge Gap examines the stability
   of an $n$-electron state with respect to
   $n+1$ and $n-1$ many-electron states and is defined as
   $$ {\rm Charge~ Gap} = \Delta_{c} = E(n+1)+E(n-1) - 2E(n).   $$ 
   The energies $E(n)$ in the above equation represent the ground state energies of the many-electron system at specified electron 
   fillings.
    Normalized double occupancies (summed over all the sites) or the average number of doubly occupied sites  in a given state of the cluster is given by $${\rm Double~ Occupations} =\sum_i w_{i}<d_{i}^{\dagger}
d_{i}>~ {\rm with}~ d_{i}^{\dagger}=c_{i\uparrow}^{\dagger}c_{i\downarrow}^{\dagger}$$ where $i$ represents a site index and $w_i$ an associated weight.
The averages shown above can be calculated for the ground state or any higher lying excited state.

\section{2$\times$N ladders: Static Case}
We illustrate the above defined charge   pairing and double occupations using 2$\times$N ladders as an example shown in Fig.~\ref{docc_full}. Note how the number of
doubly occupied sites (Fig.~\ref{docc_full}(a)) increases to their expected values at large negative $U$ values for a 2$\times$4 ladder in the ground state (for example, at quarter filling where $n=4$, the maximum number of doubly occupied sites = 2).
We are searching for ways to induce such enhancements at positive $U$ values (see previous work on ET$_2$X; Ref.~\cite{DeSilva}).

    In Fig.~{\ref{docc_full}(b)}, where the Charge Gap at one hole off quarter-filling vs $U$ is shown, the negative gap regions indicate that at those $U$ values, an equally weighted combination of $n-1$ and $n+1$ (even numbered) electron states is more stable than the state with exactly $n$ (odd numbered) electrons in this cluster.  We identify this  as a phase separation instability.
        With increasing $U$, the behavior of the Charge Gap depends on the lattice size, geometry and many electron interactions.
   Looking at Figure ~\ref{docc_full}, we observe that for small $U$ values, the negative Charge Gap regions occur in the N = 2$\times$4, 2$\times$8 ladders at the indicated filling. For N=2$\times$8 near quarter filling, it is energetically favorable for the
   mixtures of paired $n=8$ and $n=6$ electron states to coexist instead of the $n=7$ electron state
   However, note that for N=2$\times$6 and 2$\times$10 ladders, we observe
   the opposite behavior at small $U$ values (i.e., absence of phase separation as defined above). Even this behavior has
   its advantages since in this odd electron state, it is easy to induce a magnetization by applying a small, external magnetic field.
   When the paired states (with an even number of electrons) are stable, these could be classified into states where the number of doubly occupied sites is relatively high (Paired Electron Crystal (PEC)~\cite{DeSilva,clay2019}) or clusters with antiferromagnetically  coupled electrons (with a majority of singly occupied sites having up or down spins). The conclusion here is that if there is a possibility of tuning the $U$ parameter in some fashion, it is possible to induce certain emergent states with desirable magnetic or paired states, as described in the following sections. 
   
   \begin{figure}
\includegraphics[width=20pc]{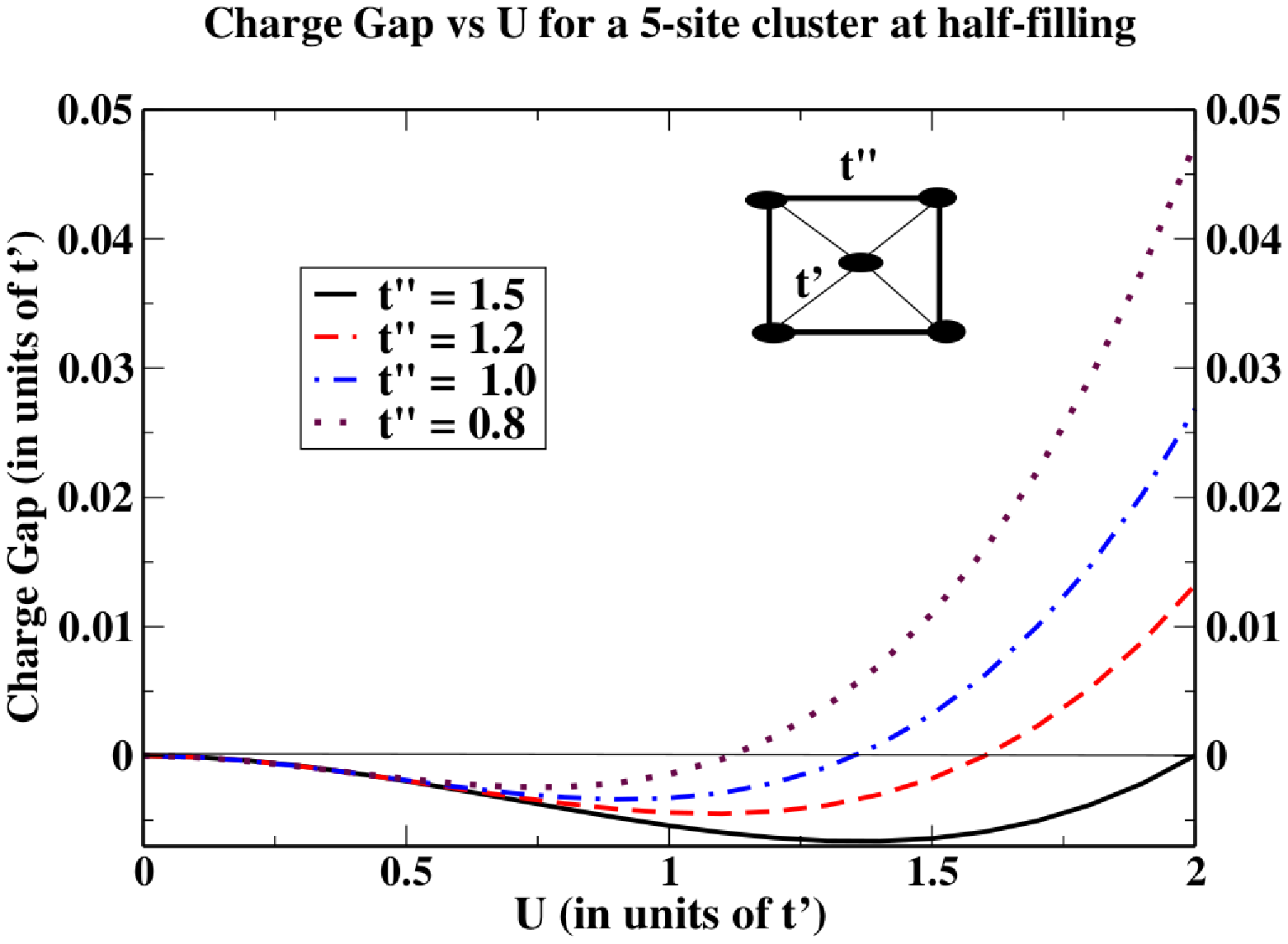}
\caption{Charge Gap at half-filling for a 5-site cluster for a single orbital Hubbard model for various values of $t^"$ (nearest neighbor hopping parameter among the outer sites).
Energies are measured in units of $t^{'}$ (hopping parameter between the center and outer sites) 
and open boundary conditions have been imposed here. 
Note that the negative Charge Gap regions imply an electronic phase separation where mixtures of many-body states with electron counts  $n=4$ and $6$ can coexist instead of $n=5.$ This geometry
closely resembles the $\kappa$-phase molecular units in CuO$_4$ and in the organic compound (BEDT-TTF)$_2$I$_3$.\label{5s-cgap}}
\end{figure}

\begin{figure}
\includegraphics[width=20pc]{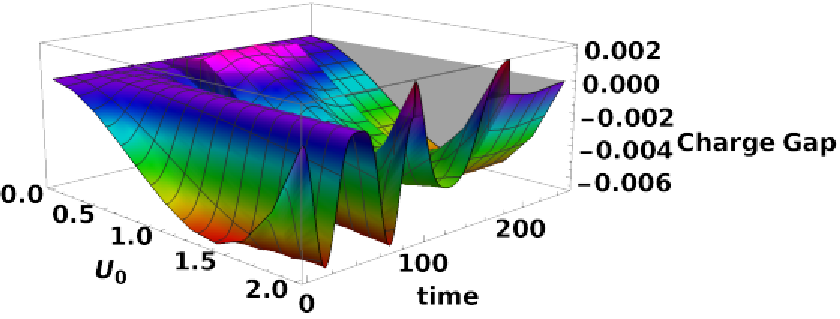}
	\caption{ The time dependence of the Charge Gap (at $t^{"}= 1.5$) under a laser pulse (as described in Equations~\ref{eqn:pulse} and \ref{hamil0}) in the 5-site cluster. The driving frequency $\omega=2.5$ in units of hopping.} Note that the laser pulse can drive the phase separation instability at $U$ values as large as 2 with the help of a small, external magnetic field $h$, as in Eq.~\ref{hamil0}.
	\label{5site:time-dep}

\includegraphics[width=20pc]{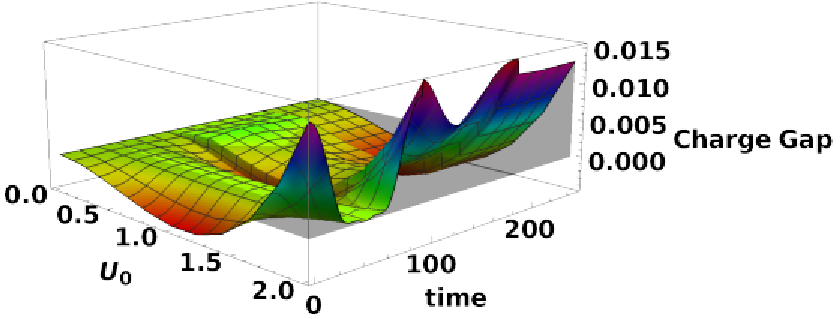}
	\caption{
	When a time dependence is introduced to the $t^{"}$ parameter related to hopping between outer sites of the 5-site cluster, as indicated in the text (Eq.~\ref{eqn:pulse2} with A=1.2), some
	of the negative Charge Gap regions in Fig.~\ref{5site:time-dep} are replaced by positive gap regions which may be identified as a consequence of counter-adiabatic driving.
	A small magnetic field $h$ (in Eq.~\ref{hamil0}) can be used to detect magnetic polarizations evident in the positive gap regions at half-filling.}
	\label{5site:time-dep1}
\end{figure}
   
   \section{Hubbard Clusters: Dynamics}  
    The exact many-body solutions of the (static) 5-site cluster with open boundary conditions show a high degree of sensitivity of the negative Charge Gap regions as a function of $U$, for small $U$ values (see Fig.~\ref{5s-cgap}). This provides an ideal platform for a simple enough case study of dynamics. This cluster closely
    resembles a CuO$_4$ unit as well as
     the $\kappa$-phase molecular unit in the organic compound (BEDT-TTF)-compounds~\cite{DeSilva}.  The Charge Gap
    at certain fillings such as half-filling and below, remains negative indicating possible phase separation and the coexistence of paired states.
    In order for this to happen, the onsite interaction parameter $U_0$ has to be less than 2 (in units of $t^{'}$) but it also depends crucially on the
    hopping parameter $t^{''}$ between the nearest neighbor sites on the boundary (such as between the O atoms in CuO$_4$). Near half-filling, suppose this cluster is driven by
    a time dependent pulse at frequency $\omega$ (as described in the experimental study of Ref.~\cite{Cavalleri}) 
    with time {\sl independent} (constant) hopping $t_{ij} = 1.5$ between outer sites (in units of $t^{'}$ as in Fig.~\ref{5s-cgap}). 
    
    Using exact diagonalization at every time step, we diagonalize the Hamiltonian
     \begin{equation}
       \mathcal{H}(t) = -\sum_{<ij>\sigma} t_{ij} B_{ij\sigma}  + 
       \sum_{i} U(t){n}_{i\uparrow}{n}_{i\downarrow}  +
       h\sum_{i}(n_{i\uparrow} - n_{i\downarrow})
       \label{hamil0}
\end{equation} 
with
    \begin{equation}
    U(t) = U_0\{ 1- \sin^2({\omega t})\exp({-\frac{(t-t_0)^2}{(2t_0)^2}}\}.
    \label{eqn:pulse}
    \end{equation}
    
     By monitoring the Charge Gap as a function of time and $U_0$, we see that the negative regions of the gap can be sustained until the pulse vanishes (see Fig.~\ref{5site:time-dep}).
    In addition, a small, external magnetic field $h$ has been added in order to keep track of possible magnetic
    polarization. Near half-filling, in the negative gap regions, the magnetization remains null while positive gap regions show a nonzero magnetization (due to an unpaired electron).
    Also, as soon as a time variation of the hopping parameter $t^{''}_{ij}$ on the boundary of the cluster similar to Eq.~\ref{eqn:pulse} is introduced, the Charge Gap
    shows positive regions for
    $0.8 \leq A\leq 1.4$ where
    
      \begin{equation}
    t^{''}_{ij}(t) = A\{ 1- \sin^2({\omega t})\exp({-\frac{(t-t_0)^2}{(2t_0)^2}}\}.
    \label{eqn:pulse2}
    \end{equation}
    This  behavior (shown in Fig.~\ref{5site:time-dep1} for $A=1.2$ in Eq.~\ref{eqn:pulse2}) may be described as being close to "counter-adiabatic" driving. It is somewhat similar to
    introducing some form of non-inertial effects in a classical system, thereby forcing it into a different path.
    The corresponding double occupations remain high at intermittent times indicating the presence of a PEC.
     For large $U_0$, it is energetically unfavorable to have doubly occupied sites (in the ground state)  due to strong Coulomb repulsion and hence it favors an Antiferromagnetic (AFM) state. In addition, the relevant moderate $U_0$ values below 8 (in units of $t^{'}$) and strong frustrations (i.e., large $t^{"}> 0.8$) are also seen to lead to a
    possible AFM $\rightarrow$ PEC transition (see also Refs.~\cite{DeSilva, clay2019}). By appropriately tuning the parameters in this problem, the resulting electronic structure and corresponding (paired vs magnetic) phases can be manipulated easily.

\begin{figure}
\includegraphics[width=16pc]{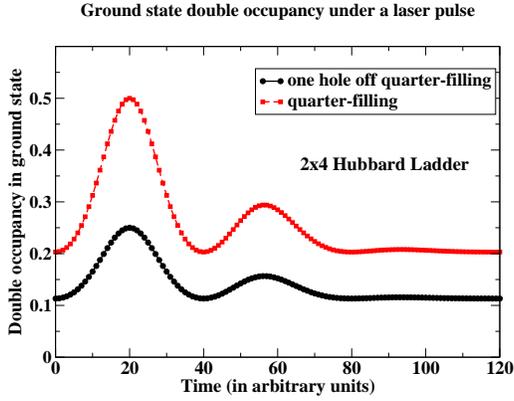}
\caption{The calculated double occupancies (averaged doubly occupied sites) of the ground state vs time $t$ at two fillings of the 2x4 ladder subjected to
an external laser pulse leading to excitations with Hubbard parameter $U$ (as in Eq.~\ref{eqn:pulse} with $U_0=3$ and $\omega=3.3$ in units of hopping). Note the enhancements beyond the ($t=0$) equilibrium values at $t=20$.  Such enhancements in double occupations are likely to favor superconductivity, as observed by Buzzi et al.~\cite{Cavalleri} for the organic material (see text). The flattening behavior of the double occupancy at large times is due to the decay of the laser pulse.}
\label{docc}
\end{figure}

\begin{figure}
\includegraphics[width=18pc]{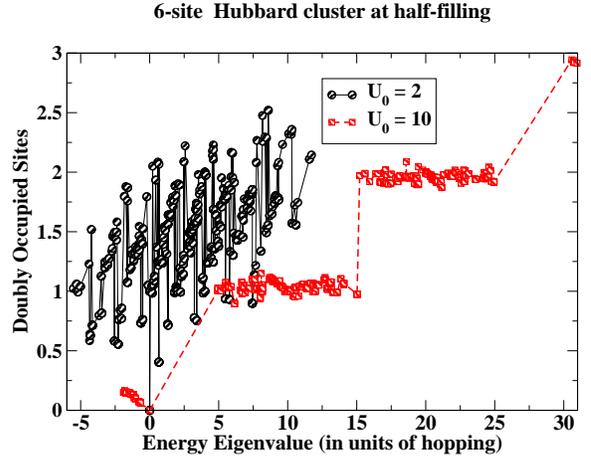}
\caption{The calculated double occupancies (doubly occupied sites) as a function of the energy eigenvalues (from the ground state to higher lying excited states) at  $U_0 =2$
and $U_0=10$  for a 6-site (static) Hubbard cluster at half-filling. Note the plateaus near integral values for $U_0=10$ (i.e., large positive Hubbard $U$ parameter). Similar clusters of eigenvalues exist for $U_0=2$ as well in  narrow bands of energy. Note that in general, higher lying eigenstates have higher double occupancies and using dipole-type excitation ideas discussed in the text, it may be possible to access those states.}
\label{doc1}
\end{figure}

   \section{Optical excitations}
   Using mid-infrared pulses, modulation of the Hubbard $U$ parameter in an organic Mott insulator was reported previously in Ref.~\cite{singla}. Recently,
   excitation of local (C=C) molecular vibrations in the CT salt $\kappa$-ET$_2$Cu[N(CN)$_2$]Br have also been
   observed to result in a large increase in carrier mobility and the opening of a superconducting-like optical
   gap ~\cite{Cavalleri}. Apparently, a coherence temperature T$^{*}$= 50K, far higher than the equilibrium
   superconducting transition temperature T$_C$= 12.5K, has been detected.
   In order to model this dynamical behavior, 
   we have monitored the double occupancies in a 8-site (2$\times$4) Hubbard ladder, 6- and 5-site clusters  using the time dependent Hubbard Hamiltonian given by Eq.~\ref{hamil0} at different fillings.
 The parameters used in these
calculations were obtained from the work of Buzzi et al.~\cite{Cavalleri} 
(for laser-induced modulation of the Hubbard $U$ parameter).

A plot of normalized double occupancies (summed over all the sites at $U_0)=3$)  in the ground state of the 2x4 ladder, $\sum_i w_{i} <d_{i}^{\dagger}
d_{i}>$ (as discussed earlier), is also shown in Fig.~\ref{docc}. Note how the double occupancies show a transient increase from their
equilibrium ($t=0$) value to that at $t=20$.
Such enhancements in double occupations (seen in other clusters as well) are likely to correlate with increases in mobility and may favor superconductivity, as observed by Buzzi et al.~\cite{Cavalleri}.

\section{Dipole-type excitations of doubly occupied states of Hubbard clusters}

This section is focused on manipulating double occupancies in Hubbard clusters using methods from Quantum Optics. Based on the many-orbital, time-dependent Schrödinger equation following Ref.~\cite{dipole}, we
demonstrate that it is possible to reach any desirable doubly occupied configuration by a sequence of laser pulses with a specific frequency $\omega$, and duration $\tau$. The time-dependent Hamiltonian,
$H(t)  = H_0 + H_1(t)$ is given by
\begin{equation}
       \mathcal{H}_1(t) = -e \cos(\omega t)\{\mathbf{\mathcal{\vec E}_{el}}\cdot\  
       \sum_{s=1}^{n} {\vec r}_s\}  
       \label{hamil2}
\end{equation} 

The electric field component here drives the inter-site electronic transitions between two many-electron states $\ket{i}$ and $\ket{j}$ 
of the unperturbed Hamiltonian $H_0$ as defined
previously. The matrix of the dipole moments
$d_{ij}$ or the Rabi frequency matrix $$\Omega_{R}^{ij}
= |d_{ij}|{\mathcal{E}_{el}}/\hbar$$ drives the time evolution of the Schrödinger wave function.

Here the calculated well-defined many-body energy levels,
$\{E_j\}$, of the cluster along with the dipole moment transition matrix, $\{d_{i,j}\}$, are used. The transitions between the ground state, $g$, and the states with high double occupancies (Fig.~\ref{doc1}), $j=doc$, driven by the appropriate dipole moment, ${d}_{g,doc}$, are excited by the laser pulse having transition frequency, $\omega_{g,doc}=(E_{doc} - E_g)/\hbar$. This coherent excitation of all initially
populated ground states into states with high double occupancy requires a finite-length laser pulse with
duration $\tau=\pi/\Omega_{g,doc}$, where
the characteristic frequency $\Omega_{g,doc}=\sqrt{{\Omega_R}^2 +\omega_{g,doc}^2}$.
Here the Rabi frequency, $\Omega_R=d_{g,doc}E_0/\hbar$, depends on both the amplitude $E_0$ of the electric field of the laser, and the dipole moment, $d_{g,doc}$, between the ground and the doubly occupied state.
Our numerical estimate of $d_{g,doc}\simeq 5er$
($r\simeq$ being a nearest neighbor distance and $e$ = electronic charge) yields a frequency $\Omega_{g,doc}$,
with the required duration of a pulse, $\tau\simeq 2-5$ fs.
One can see that all required parameters of laser excitation favorable to high double occupancy  are within the accessible range.
Based on the numerical results from our cluster studies (see Fig.~\ref{doc1}), the energy difference is $\Delta E_{g,doc}\simeq 1.5 U$. Since the typical $U\simeq 2-3 eV = 2-3t$, the coherent excitation of doubly occupied states requires a laser within the visible-near-infrared optical range.

        In summary,
         small condensed matter systems that are driven can give rise to various emergent phases.
         Floquet-Bloch effects can lead to quasi-energy states resulting in gap closures and other unexpected electronic properties. In small Hubbard clusters, at appropriate $U$ values, it is possible to tune phase separation, dipolar or magnetic-type instabilities either by variation of filling, temperature and time dependent perturbations such as femto-second laser pulses. 
        Some of these calculations can provide insights into the electronic properties of certain superconductors, including phase separation and electromagnetically induced emergent phases.

     \acknowledgements
      We acknowledge the computing resources provided by the Center for Functional Nanomaterials, which is a U.S. DOE Office of Science Facility, at Brookhaven National Laboratory under Contract No. DE-SC0012704 and  the Swedish national infrastructure for computing - SNIC provided by the high performance computing center north (HPC2N). We acknowledge support from a Los Alamos grant through the U.S. DOE. This research also used funding from the ERC synergy grant "HERO" (No 810451), VILLUM FONDEN via the Centre of Excellence for Dirac Materials (Grant No. 11744), the Knut and Alice Wallenberg foundation ( 2019.0068) as well as the Vetenskapsrådet (no. 2017.03997). \revision{We thank A. V. Balatsky and R. T. Clay for helpful discussions and A. Cavalleri for a preprint of their recent work. One of us (GWF) gratefully acknowledges hospitality and discussions with Hai-Qing Lin and colleagues at the Beijing Computational Science Research Center.}

\end{document}